# A Polymeric Planarization Strategy for Versatile Multi-terminal Electrical Transport Studies on Small, Bulk Quantum Materials

Huandong Chen[1], Amir Avishai[2], Jayakanth Ravichandran[1,2,3*]

[1]Mork Family of Department of Chemical Engineering and Materials Science, University of Southern California, Los Angeles, California, USA

[2]Core Center for Excellence in Nano Imaging, University of Southern California, Los Angeles, California, USA

[3]Ming Hsieh Department of Electrical and Computer Engineering, University of Southern California, Los Angeles, California, USA

*e-mail: j.ravichandran@usc.edu

**Abstract**

We report a device fabrication strategy of making multi-terminal electrical contacts on small (< 1 mm) bulk quantum materials using lithography-based techniques for electrical transport studies. The crystals are embedded in a polymeric medium to planarize the top surface, and then standard lithography and microfabrication techniques are directly applied to form electrodes with various geometries. This approach overcomes the limitations of crystal thickness and lateral dimensions on establishing electrical contacts. We use low stress polymers to minimize the extrinsic thermal strain effect at low temperatures, which allow reliable transport measurements on quantum materials that are sensitive to strain. The crystal surface planarization method has enabled electronic transport studies such as in-plane anisotropy, Hall measurements on small, bulk $BaTiS_3$ (BTS) crystals, and provides unique opportunities for two-dimensional (2D) heterogeneous integration on three-dimensional (3D) / quasi-one-dimensional (quasi-1D) bulk materials. Our strategy is general for many small, non-exfoliable crystals of newly synthesized quantum materials and paves the way for performing versatile transport studies on those novel materials.

**Introduction**

Electrical transport measurements on quantum materials not only play an important role in understanding the nature of electronic phases[2-4] and identifying phase transitions[5-7], but also provide supporting experimental evidence for phenomena such as symmetry breaking[8-10]. For many decades, epitaxial semiconductor thin films and heterostructures such as GaAs/AlGaAs two-dimensional electron gas (2DEG) systems were the only few material platforms available for electrical transport studies, especially quantum transport,[11, 12] before low-dimensional nanomaterials such as carbon nanotube[13], graphene[14, 15] and transition metal dichalcogenides (TMD)[16, 17] emerged. Recent advancements in nanomaterials synthesis, mechanical exfoliation and lithography techniques over the past few decades have successfully expanded transport studies to a large class of nano-scale and exfoliable layered materials[13-17]. On the other hand, there are also a large variety of newly synthesized or theoretically predicted quantum materials that are not easily exfoliable down to lithographically compatible thicknesses such as many quasi-one-dimensional (quasi-1D) and 3D crystals[18-20], and hence, multi-terminal electrical transport studies are limited on those materials.

Manual bonding is still the most widely adopted method for making electrical contacts to those bulk crystals[20-22], where thin metal wires such as gold, platinum, or indium are directly bonded onto the crystal using conductive epoxy or self-melting methods. Sometimes a pre-sputtered Au layer through a shadowing mask is also applied to further reduce the contact resistance[23]. To make a four-probe measurement contact geometry, this manual bonding method would usually require at least mm-scale crystal size and significant user skills[21]. However, newly synthesized crystals of quantum materials from either chemical vapor transport (CVT) or flux growth methods are typically small and in a range of a few hundred micrometers. Extensive growth condition optimizations are required to achieve crystals larger than a millimeter in lateral dimensions[24, 25]. Even for quasi-1D morphology crystals that are long enough for standard four-probe geometry, achieving more complicated electrode designs such as Hall bar geometry and angular-dependent electrodes is still challenging due to geometrical limitations on other directions[22]. Another option is to use a prefabricated $SiN_x$ shadowing mask[26, 27], but electrode patterning accuracy, alignment, and crystal handling remain critical limitations.

Currently, most of the advanced multi-terminal electrical transport studies on bulk single crystals rely on sophisticated sample preparation procedures such as focused ion beam (FIB) micro-patterning, where a lamella with a few micrometers thick is typically lifted out from a bulk crystal and then carved into desired patterns, such as Hall bar geometry, in order to study its charge transport behavior[28-33]. Using this technique, a number of intriguing quantum systems, including iron pnictides[28, 29], semimetals[30, 31], and heavy-fermion superconductors[32, 33], have been fabricated into a variety of microstructures for transport studies, whereas only a few research groups mastered and frequently used it. The complexity of the lamella lift-out procedure, which requires extensive TEM sample preparation experiences, is the primary factor limiting the extensive applications of this method.

On the other hand, the simple idea of crystal embedding using epoxy resin has been applied to handle small bulk samples in situations such as microtome sectioning[34, 35] and crystal surface polishing for chemical analysis[36], but its applications towards electronic devices fabrication or electrical transport measurements have not been explored extensively. Recently, Kang *et al* reported the fabrication of lithography-defined GaAs photoelectrodes[37] by embedding freestanding microcells (500 μm in lateral dimension and 4-5 μm thick) in UV-curable epoxy NOA. However, these epoxies typically exhibit relatively large CTE (coefficient of thermal expansion,

~ 80 ppm/K for NOA 61), and a significant amount of thermal strain (~ 1.6% at 100 K for NOA) could be introduced at low temperatures. To obtain reliable and reproducible transport measurements on quantum material systems that typically have strong electron-lattice interactions and are sensitive to strain fields, low-stress embedding medium is needed.

In this article, we adopt the polymeric embedding approach and demonstrate a simple crystal planarization strategy that uses a low-stress polymer as the embedding medium, which allows reliable lithography-compatible multi-terminal electrical transport studies on small, bulk single crystals. We have chosen the quasi-1D hexagonal chalcogenide $BaTiS_3$ as a model non-exfoliable quantum material system for this demonstration. Recently, a subset of the authors reported two unique electronic phase transitions and charge density wave (CDW) ordering in BTS at low temperatures using this approach[38]. By adopting the planarization strategy, we characterize in-plane conductivity anisotropy of $BaTiS_3$ from orientation-resolved four-probe measurements. We also demonstrate the direct microstructural patterning of Hall bars on pre-embedded crystals using plasma ion milling, and present detailed Hall measurements on micro-patterned Hall bar device. In addition, by integrating transferred electrodes on bulk crystals, we show that the planarized platform provides unique opportunities towards 2D/3D integration, which may lead to a completely new region of rich physics and device applications. Our method is generally applicable to many other non-exfoliable quantum systems and paves the way for advanced transport measurements on those novel systems that were previously impractical.

## Results and Discussion

**Crystal planarization using polyimide as an embedding medium**

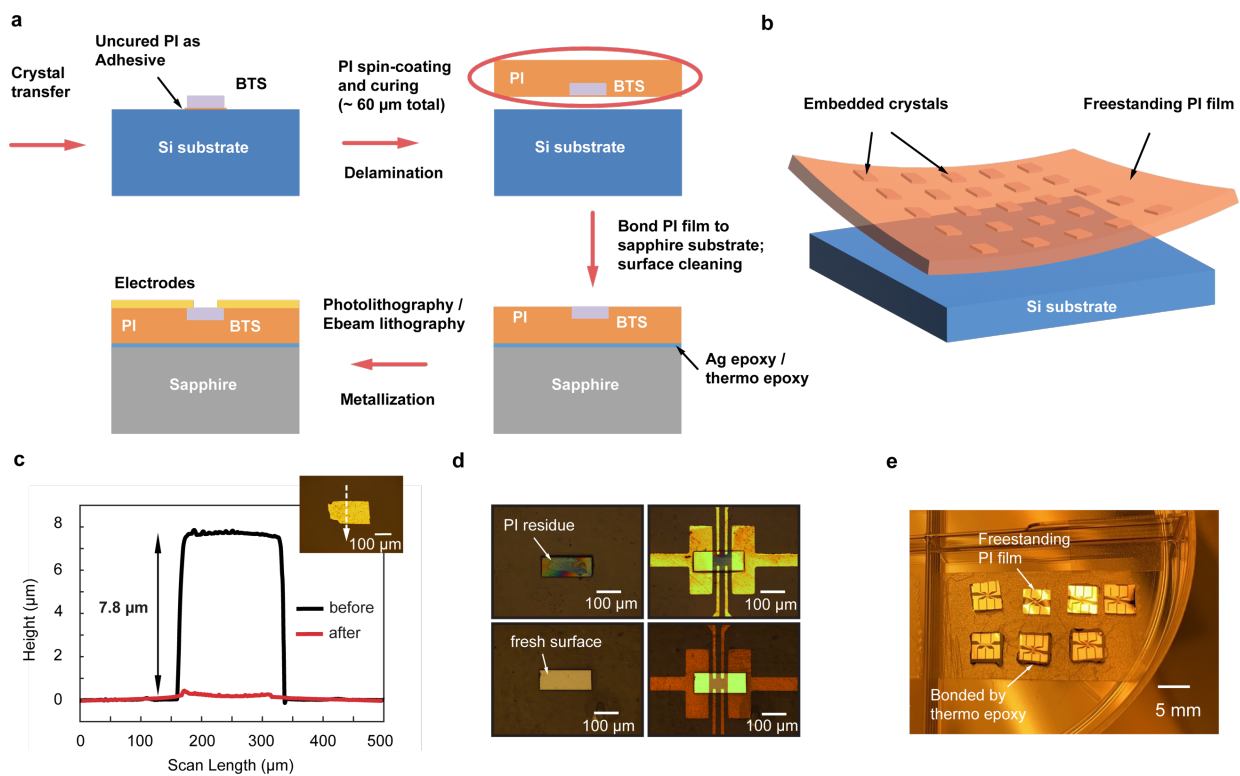

**Figure 1. a**, Schematics of the fabrication process flow for a bulk BaTiS$_3$ device utilizing PI as the embedding medium. **b**, Schematic illustration of PI peeling off step with 5 × 5 crystals array processed at the same time. **c**, Surface profile scans of an as-grown BaTiS$_3$ platelet crystal placed on a PDMS stamp and the same crystal after polymeric embedding. **d**, Optical micrographs of BaTiS$_3$ crystal after crystal embedding (top left), surface cleaning (bottom left), metallization with (top right) and without (bottom right) thin PI dielectric on channel region. **e**, Photographic image of multiple BTS devices fabricated for transport studies.

Single crystals of BTS were synthesized using a chemical vapor transport method[39], and as grown 001-oriented platelet crystals with 5-20 μm thick[40] and 100-200 μm wide were selected for the studies discussed here. Such sizes are typical for early-stage bulk quantum materials synthesis using CVT or flux growth, particularly before extensive optimization efforts on the growth conditions are applied[24, 25]. Unlike many 2D materials assembled by weak van der Waals interactions, the interchain ionic bonding between TiS$_6$ octahedra chains and Ba chains of BTS makes it difficult to achieve atomically thin flakes with large lateral dimensions by mechanical exfoliation[39]. Hence, direct application of lithography techniques is challenging due to the substantial thicknesses of the crystals. Moreover, because of the limited bulk crystal lateral sizes, conventional contacting methods such as manual bonding are challenging for BTS, too.

Figure 1a illustrates a schematic of the device fabrication process flow for embedding a bulk BTS crystal in a low-CTE polyimide (~ 3 ppm/K) medium for electrical transport measurements. We first place the as-grown crystal on a flat Si wafer using a polydimethylsiloxane (PDMS) elastomer stamp attached to a standard transfer stage. To temporarily bond the crystal to the substrate, a thin layer of uncured polyimide (PI) is used as an adhesive, followed by a partial curing step to fix the crystal position before moving on to the next crystal. Multiple crystals can be processed on the same wafer by repeating crystal transfer and releasing steps, as shown in Figure 1b. We then spin-coat and cure the same PI precursors multiple times to completely cover the crystals. A typical polyimide film thickness of about 60 μm provides sufficient mechanical support and is suitable for crystals with thicknesses below 30 μm. For thicker crystals, the conformal coating of PI may become an issue and lead to a lower yield. Freestanding PI films are cut and peeled off from the Si substate, leaving the top surface of BTS crystal planarized. After that, the PI chip is flipped and attached to a sapphire substrate with thermally conductive epoxy or tape.

Surface planarization is critical in enabling successful application of lithography techniques on sub-mm bulk crystals. Here, we take advantage of the controllable adhesion strengths between PI film and the flat silicon wafer, which serves as the crystal's temporary molding surface. The step height of the crystal-substrate boundary is reduced from the as-grown crystal thickness (5 – 20 μm) to less than 200 nm after embedding, as shown by the surface profile scans in Figure 1c. Hence, lithography and thin film metal deposition can be readily performed on those bulk crystals. Moreover, compared to the conventional contacting method involving manual bonding, which restricts the choice of metals and electrode geometry, the lithography-compatible microfabrication method offers complete flexibility for contact optimization, which is particularly crucial for semimetals and semiconductors[41, 42]. The fabrication procedures can be designed and optimized based on the specific physical properties of the crystal, and the processing is largely identical to that of handling large single crystalline wafers or 2D materials-on-substrate systems. Figure 1d illustrates the optical micrographs of BTS devices at various processing steps and Figure 1e shows a photographic image of multiple devices fabricated for transport studies.

**In-plane conductivity anisotropy from orientation-resolved four-probe measurements**

One of the biggest advantages of this planarization strategy is its compatibility with lithography and other standard microfabrication techniques. For instance, versatile electrode geometries and experiments as required can be designed to probe the underlying charge transport behavior of those non-exfoliable bulk systems with limited sizes. We have previously demonstrated the observation of multiple phase transitions in BTS from transport measurements along the chain axis (*c*-axis) using the same PI embedding method, and we have confirmed its validity and reliability by comparing the transport behavior with that from conventional manual bonding method and between repeated measurement scans, respectively[38]. Here, we further characterize the in-plane anisotropic transport behavior of BTS from orientation-resolved four-probe measurements, where a total of 8 electrodes were used.

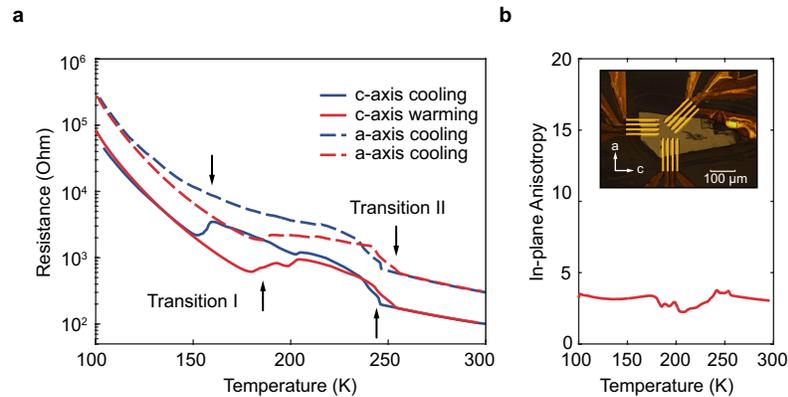

**Figure 2. a**, Four-probe electrical resistance of $BaTiS_3$ along the *c*-axis and *a*-axis as a function of temperature. The transport data was collected from the same $BaTiS_3$ device with linear four-probe electrodes along different crystal orientations, as shown in the inset of Fig. 2b. **b**, Temperature dependence of the in-plane conductivity anisotropy (defined as $R_a/R_c$). The anisotropy value is between 3 to 4 throughout the measurement range with anomaly showing near transitions.

Electrical conductivity anisotropy, a direct measure of crystal structural anisotropy and electronic anisotropy, is one of the most important characteristics of quantum materials[23, 43] such as BTS, and it can be useful in understanding the underlying physics behind CDW transitions and for developing novel anisotropic electronic and optoelectronic devices[44, 45]. Figure 2a plots the temperature dependent resistance of a BTS platelet along both *c*- and *a*-axes, and the microscopic image of the device is shown as the inset of Figure 2b. The in-plane conductivity anisotropy of BTS is obtained by directly calculating the ratio of resistance along *a*- and *c* directions ($R_a/R_c$) and assuming the sample thickness is uniform. At room temperature, this value is close to 3, which is not very large compared with many other conventional quasi-1D CDW systems. Upon cooling, the resistance along both directions increases and BTS undergoes a CDW transition near 240 K that we refer to as Transition II. Upon further cooling, BTS undergoes another phase transition at 150 K (Transition I) with a sharp drop of resistance along the *c*-axis that is consistent with the reported transport behavior on a needle-like crystal along *c*-axis[38], while the resistance measured along a-axis reveals a more gradual transition into the low-temperature phase, which could be associated with the unique CDW ordering observed in the *a-b* plane. The anisotropy value varies from 3 to 4 across the entire measurement temperature range, with marked anomalies at the transitions. Hence, the polymeric planarization strategy has enabled direct and accurate in-plane electrical anisotropy determination even on small bulk crystals.

**Hall bar structure patterning from pre-embedded crystal using PFIB milling**

Multi-terminal Hall bar structure is usually preferred in magneto-transport measurements to accurately and simultaneously determine transport properties such as carrier concentration, mobility, and magnetoresistance[16]. Hall bar structures are usually formed by lithography and dry etching for thin film or flakes of 2D materials[16, 46], whereas the processing is much harder for sub-mm sized bulk single crystals. FIB microstructure patterning is one of the most well-established and advanced routes to handle those non-exfoliable bulk crystals as stated earlier[29, 30]. However, the lamella preparation for milling is complicated and has limited wide applications of this method, whereas the milling process on the lamella itself is relatively straightforward and quick. As shown in Figure 3a, we demonstrate a one-step fabrication procedure for forming multi-terminal Hall bar microstructures of BTS directly from a prefabricated device with desired metal contacts, using plasma FIB (PFIB). The pre-patterned metal contacts not only help in mitigating the electron beam charging issues during processing, but also provide unique opportunities to investigate the effects of microstructures and ion implantations on transport behavior by comparing measurements taken before and after patterning. The Xe plasma milling process was performed to form the desired geometry, followed by a fine polishing step to clean up the damaged crystal surface. To ensure the electrical connection and maximize the contact area, an additional layer of Tungsten (W) was directly deposited on top of the contact region by PFIB.

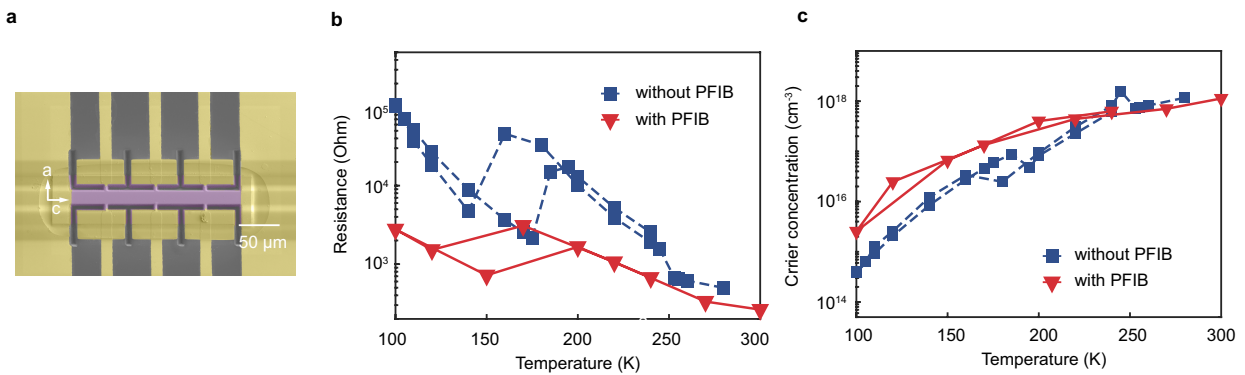

**Figure 3. a**, False colored scanning electron micrograph of Hall bar structures fabricated using PFIB milling on pre-embedded BaTiS$_3$ crystals with metal contacts for magneto-transport measurements. **b-c**, Temperature dependent resistance and carrier concentration of BTS devices with and without PFIB microstructure patterning. Part of the data of 'without PFIB' condition was reproduced from Ref 38.

Hall measurements from PFIB-fabricated devices are shown in Figure 3b and 3c. Compared with the measurements on devices without FIB micropatterning, the resistance level is lower, whereas both phase transitions are maintained. The reduced resistance is mainly attributed to the increased carrier concentrations in the system, which potentially comes from the FIB-induced damage on the channel crystals during the milling processes. If the damage is from the surface layer, a final surface polishing step with a lower milling voltage could help reduce any extrinsic surface contributions to transport.

**Integration of transferred electrodes on BaTiS$_3$ bulk crystal**

In addition to the intrinsic physical properties of the quantum material itself, heterogeneous integration with other interesting systems opens new opportunities for studying rich physics from proximity effects[47]. The emergence of heterostructures of 2D materials, including the twistronics, has offered unique platform and great flexibility for studying interesting physics such as superconductivity[46, 48] and correlated insulating states[49, 50]. In principle, the crystal surface planarization strategy presented here also provides opportunities for heterogeneous integration of 2D materials such as graphene and transition metal dichalcogenides on various bulk quantum materials using regular transfer techniques. Here, we use 50 nm thick Au electrodes as an example and show their integration on BTS crystal using the transfer processes modified from the procedures previously reported[42, 51]. Au electrodes with desired geometries were prepared on a clean silicon wafer using standard photolithography and ebeam evaporation. We use photoresist/PPC/PDMS stack to pick up and transfer the electrodes onto planarized BTS crystals, as illustrated in Figure 4a. Two-terminal I-V characterization was shown in Figure 2b, which indicates a conformal contact between Au electrodes and BTS crystal surface and sheds light on future integration of various 2D materials on bulk crystals.

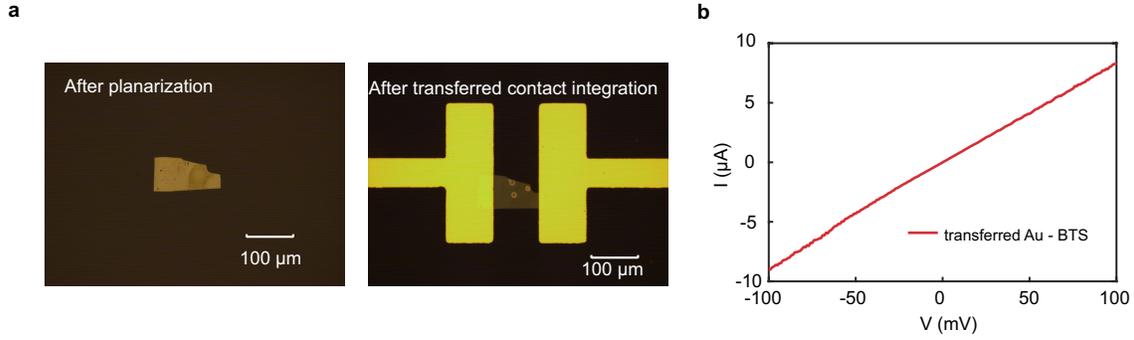

**Figure 4. a**, Optical microscopic images of BTS crystal after surface planarization and van der Waals contact integration. **b**, current-voltage characteristics of a BTS device contacted by 50 nm transferred Au electrodes.

## Conclusions

In conclusion, we have demonstrated a simple and general crystal surface planarization strategy for making multi-terminal electrical contacts on small, bulk samples with standard lithography techniques, which can be readily used for multi-terminal transport measurements including anisotropy, magneto-transport, photocurrent, *etc*. The planarization strategy also provides unique opportunities for integrating 2D materials onto bulk quantum systems. The PI-based processing conditions presented here are optimal for strain-sensitive quantum materials and work best for crystals with thicknesses up to 30 μm. The lateral size is limited only by the lithography technique to be used; hence, bulk crystals with lateral dimensions even below 100 µm can be easily handled. Moreover, we anticipate this method will be broadly useful for transport studies of *in situ* strain engineering as well, where the freestanding PI film with embedded crystal (3D/quasi-1D, or even 2D) and pre-patterned electrodes is mechanically stretched. The on-chip integration of metal wire strain gauge becomes possible and can further accurately determine the strain applied onto the material.

## Author contributions

H.C. conceived the idea and designed the experiments. J.R. supervised the project. H.C. fabricated the devices and performed electrical transport measurements. A.A. and H.C. fabricated the Hall bar microstructure using PFIB. H.C. and J.R. wrote the manuscript with input from all other authors.

## Notes

The authors declare no competing financial interests.

## Acknowledgements

This work was supported by an ARO MURI with award number W911NF-21-1-0327, an ARO grant with award number W911NF-19-1-0137 and an NSF grant with award number DMR-2122071. The authors gratefully acknowledge the use of facilities at John O'Brien Nanofabrication Laboratory and Core Center for Excellence in Nano Imaging at University of Southern California for the results reported in this manuscript.

# Supplemental Information for

# A Polymeric Planarization Strategy for Versatile Multi-terminal Electrical Transport Studies on Small, Bulk Quantum Materials

**Authors:** Huandong Chen[1], Amir Avishai[2], Jayakanth Ravichandran[1,2,3]*

[1]Mork Family of Departments of Chemical Engineering and Materials Science, University of Southern California, Los Angeles, California, USA

[2]Core Center for Excellence in Nano Imaging, University of Southern California, Los Angeles, California, USA

[3]Ming Hsieh Department of Electrical and Computer Engineering, University of Southern California, Los Angeles, California, USA

*e-mail: j.ravichandran@usc.edu

## Supplementary Methods

**Fabrication of Bulk BaTiS$_3$ Device with Polyimide**

The crystal thickness of a BTS platelet is typically 5-20 µm. The preparation of bulk BTS device started by attaching the picked crystal to a PDMS elastomer stamp (Sylgard 184, Dow Corning; precursor:curing agent = 10:1, by weight, cured at 65°C for 1 hr), which was later loaded onto a home-built transfer stage. To temporarily bond the crystal to the substrate, a thin layer of uncured polyimide (PI 2611, HD Microsystems) was applied on a pristine Si wafer (4 cm × 4 cm) and spread as adhesive using a Q-tip (TX751B, Texwipe). We roughly aligned the crystal to the substrate under a stereo microscope and then lowered stage till the crystal in contact with the PI adhesive. After holding for ~ 30s, we slowly lifted the stage, leaving a BTS crystal printed on the Si substrate. A partial curing step at 120°C for 5 minutes was applied to fix the crystal in position before moving on to the next crystal. Typically, multiple crystals up to 7 × 7 arrays are prepared on the same Si wafer to save total processing time. A similar transfer method using PI adhesive has been adopted to demonstrate heterogeneous integration of GaAs field-effect transistor (FET) arrays onto polyimide-coated glass substrate[52]. We then spin-coated the same PI precursors consecutively for four times at 1500 rpm to completely cover the crystals, with a soft-bake step at 170°C for 5 min (5°C/min ramp rate) in between and a final curing step at 200°C for 15 hours. Razor blade and tweezers were used to gently cut and peel off the PI film (6 mm × 6 mm) from the Si substrate, leaving a freestanding PI chip with the top surface of BTS crystal planarized. After that, the PI chip was flipped and attached to a 5 mm × 5 mm sapphire substrate with two-part thermally conductive epoxy (Thermo-bond 180) or silver epoxy (EPO-TEK H20E). Double-sided thermal release tape can also be used for temporary bonding to go through lithography and metal deposition processes, resulting in a BTS device on a freestanding PI chip with metal contacts. It is worth noting that no silane-based adhesion promoter (VM 651, HD MicroSystems) was applied prior to PI spin-coating for a higher peeling-off yield. All the processing steps including transfer and PI coating were performed in a fume hood environment with good ventilation to avoid inhalation of n-methyl-2-pyrollidone (NMP), which is the main solvent of PI precursor.

For metallization, a negative photoresist (AZ nLOF-2070, Merck KGaA) was spun-coated directly on polyimide chip at 4000 rpm, yielding a 6 μm flat film that was then UV-exposed (240 mJ/cm$^2$, MJB3) and developed (2 min, AZ 726 MIF) to form the electrode patterns. Fresh crystal surfaces in contacting areas were exposed by removing thin PI residues (< 100 nm) with RIE treatment (O$_2$/CF$_4$ = 45/5 sccm, 100 W, 100 mTorr, 1 min, Oxford PlasmaPro 80), leaving the rest of the crystal surface covered by the PI dielectric. Prior to depositing Ti/Au (3/300 nm) via ebeam evaporation (Temescal), the device was treated with SF$_6$ RIE (Ar/SF$_6$ = 50/15 sccm, 100 W, 200 mTorr, 1 min) and diluted NH$_4$OH (NH$_4$OH:DI = 1:10, 1 min), both of which help in surface oxide removal. After metallization, the lift-off process was carried out by immersing the device in acetone for 5 min. The device was further annealed at 170˚C for 2 min in a glovebox filled with nitrogen can be added to further improve the contact quality.

**Electrical Characterization**

Transport measurements were carried out in a JANIS 10 K closed-cycle cryostat from 100 K to 300 K. Standard low-frequency ($f$ = 17 Hz) AC lock-in techniques (Stanford Research SR830) were used to measure sample resistance in four-probe geometry, with an excitation current of about 100 nA. In-plane conductivity anisotropy was characterized by measuring standard temperature-dependent four-probe resistance along both the c- and a-axes during cooling and warming scans.

TLM analysis was done by measuring two-probe resistance between metal pads with varying spacing distances on a BTS needle-like sample at room temperature. Hall measurements were performed in a 14 T PPMS system (Quantum Design). AC current was generated by a lock-in amplifier and passed through the device, while $I_{AC}$, $V_{xx}$ and $V_{xy}$ were recorded simultaneously.

**Fabrication of BaTiS$_3$ Microstructures using PFIB**

The micro-structuring was performed directly on a regular prefabricated BTS device with multi-terminal electrodes using a ThermoFisher Helios G4 PFIB UXe Dual Beam system. The milling process was carried out at 30 kV and 15 nA to form the desired geometries such as 'cloverleaf' and Hall bar geometry. To ensure the electrical connection at the crystal/polymer boundary and to maximize the contact area, additional layer of Tungsten (W; ~ 500 nm thick) was deposited on top of the contact region at 8 kV using MultiChem Gas Delivery System in PFIB chamber. A final fine polishing step was performed at 30 kV, 1 nA to clean up the side wall and crystal surface.

**Transferred electrodes integration on BaTiS$_3$ crystal using PR/PPC stack**

The Au electrodes transfer processes are modified from the procedures previously reported[42, 51]. We use photoresist/PPC/PDMS stack to pick up and transfer the electrodes instead of PMMA/PDMS for the ease of processing. 50 nm Au electrodes with desired geometries were first prepared on a clean silicon wafer using standard photolithography and ebeam evaporation. An hexamethyldisilazane (HMDS) vapor treatment was carried out on the silicon substrate (kept at 130°C on hotplate) for 20 min to functionalize the whole wafer. Photoresist AZ1518 was then spun coated (4000 rpm) on top of metal electrodes, followed by a spin-coating of a thin PPC layer (4000 rpm). PDMS stamp was then used to pick up the whole PPC/PR/Au stack, achieving a very high yield of more than 90% without cracking Au electrodes.

The PI-embedded BTS crystal was first gently clean by $SF_6$ RIE (Ar/$SF_6$ = 50/15 sccm, 100 W, 50 mTorr, 40 s) to expose the fresh surface. Without much delay, the PPC/PR/Au electrode stack was aligned and transferred onto the BTS crystal using a home-built transfer stage. Then the whole sample stack was treated at 120°C for 5 min to release PDMS stamp by melting PPC layer. The remaining photoresist was later removed by acetone and IPA. A short bake of 110°C for 30 s is applied to further improve the electrode adhesion.

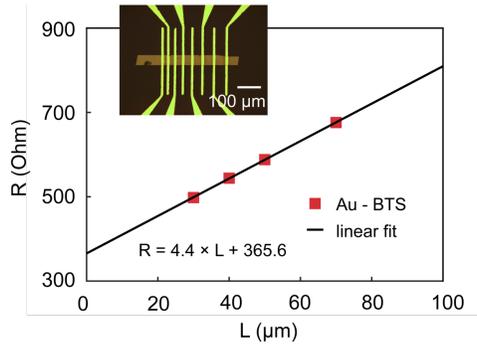

**Figure S1.** Total resistance (R) as a function of metal pad spacing (L) from standard transmission line model (TLM) measurements[1], where Ti/Au (3 nm/300 nm) were deposited on $SF_6$-cleaned $BaTiS_3$ crystal surface and annealed at 170˚C, 2 min in $N_2$ atmosphere. This specific device has an extracted contact resistance of ~ 370 Ω and a contact resistivity of $0.015\ \Omega \cdot cm^2$.